\documentclass{article}

     \PassOptionsToPackage{numbers, compress}{natbib}

    \usepackage[final]{neurips_2024}

\usepackage{graphicx}
\usepackage{subcaption}
\usepackage{float}
\usepackage{enumitem}
\usepackage[numbers]{natbib}
\setitemize{noitemsep,topsep=0pt,parsep=0pt,partopsep=0pt}
\usepackage{lipsum}
\usepackage[utf8]{inputenc} 
\usepackage[T1]{fontenc}    
\usepackage{hyperref}       
\usepackage{url}            
\usepackage{booktabs}       
\usepackage{amsfonts}       
\usepackage{nicefrac}       
\usepackage{microtype}      
\usepackage{xcolor}         
\usepackage{listings}
\usepackage{xcolor}

\definecolor{codegreen}{rgb}{0,0.6,0}
\definecolor{codegray}{rgb}{0.5,0.5,0.5}
\definecolor{codepurple}{rgb}{0.58,0,0.82}
\definecolor{backcolour}{rgb}{0.95,0.95,0.92}

\lstdefinestyle{mystyle}{
    backgroundcolor=\color{backcolour},   
    commentstyle=\color{codegreen},
    keywordstyle=\color{magenta},
    numberstyle=\tiny\color{codegray},
    stringstyle=\color{codepurple},
    basicstyle=\ttfamily\footnotesize,
    breakatwhitespace=false,         
    breaklines=true,                 
    captionpos=b,                    
    keepspaces=true,                 
    numbers=left,                    
    numbersep=5pt,                  
    showspaces=false,                
    showstringspaces=false,
    showtabs=false,                  
    tabsize=2
}
\title{Towards unearthing neglected climate innovations from scientific literature using Large Language Models}

\usepackage{hyperref}

\makeatletter
\newcommand{\printfnsymbol}[1]{%
  \textsuperscript{\@fnsymbol{#1}}%
}
\makeatother
\author{%
César Quilodrán-Casas$^{1,2\thanks{Equal contribution}}$ \quad Christopher Waite$^{1^*}$ \quad \\ \textbf{Nicole Alhadeff}$^1$ \quad \textbf{Diyona Dsouza}$^{3}$ \quad \textbf{Cathal Hughes}$^{1}$\\
\textbf{Larissa Kunstel-Tabet}$^2$ \quad \textbf{Alyssa Gilbert}$^2$ \\
$^1$ Undaunted, Grantham Institute for Climate Change and the Environment, Imperial College London\\ \quad $^2$ Department of Earth Science and Engineering, Imperial College London\\
$^3$ Dyson School of Engineering, Imperial College London\\
}

\begin{document}
\maketitle

\begin{abstract}
Climate change poses an urgent global threat, needing the rapid identification and deployment of innovative solutions. We hypothesise that many of these solutions already exist within scientific literature but remain underutilised. To address this gap, this study employs a curated dataset sourced from OpenAlex, a comprehensive repository of scientific papers. Utilising Large Language Models (LLMs), such as GPT4-o from OpenAI, we evaluate title-abstract pairs from scientific papers on seven dimensions, covering climate change mitigation potential, stage of technological development, and readiness for deployment. The outputs of the language models are then compared with human evaluations to assess their effectiveness in identifying promising yet overlooked climate innovations. Our findings suggest that these LLM-based models can effectively augment human expertise, uncovering climate solutions that are potentially impactful but with far greater speed, throughput and consistency. Here, we focused on UK-based solutions, but the workflow is region-agnostic. This work contributes to the discovery of neglected innovations in scientific literature and demonstrates the potential of AI in enhancing climate action strategies.
\end{abstract}

\section{Introduction}

The International Energy Agency (IEA) notes that about half the projected CO2 reductions that will be required to
achieve Net Zero by 2050 will depend on technologies that are currently not commercially
viable– highlighting the critical need for breakthrough innovations to mitigate the impacts
of climate change \cite{iea2021}. Unlike the clear relationship between life sciences and biotech innovation, for example, there is no one academic field that dominates climate innovation. Potential solutions can emerge from disparate fields. Therefore, one likely reason for the neglect of certain climate solutions is the sheer volume and diversity of scientific literature. Traditional methods of knowledge discovery and synthesis may fail to capture innovative approaches buried in vast datasets, leading to missed opportunities for policy and technological advancement \cite{bornmann2015growth}. This is especially relevant for countries like the UK, which has a world-leading academic culture and made substantial investments to foster climate innovation but may still have untapped potential in its existing scientific outputs \cite{BEIS2022}. To address this challenge, we propose the use of machine learning (ML) and Large Language Models (LLMs) to systematically identify climate innovations in scientific literature. We leverage OpenAlex \cite{priem2022openalex}, a comprehensive open dataset of scholarly papers and comprehensive meta-data, to provide test data for analysis by state-of-the-art language models, such as GPT4-o from OpenAI. These models are prompted to evaluate paper abstracts across seven dimensions: climate emissions reduction/removal potential, technology level, deployability, market need, potential to enable subsequent innovation, mission focus of research, and neglectedness. Our hypothesis is that the research evidence base for many high-impact climate solutions is already documented in scientific papers from the UK but these have not yet been fully identified or systematically prioritised. Benchmarking the outputs of the LLMs against parallel human evaluations, we aim to assess the effectiveness of these models in finding overlooked innovations and identify any potential advantages over human reasoning. This research contributes to both the discovery of neglected climate solutions and the application of ML in enhancing domain-specific knowledge extraction, potentially accelerating climate action by uncovering actionable insights hidden within the existing literature.
\section{Background and Literature Review}

Machine learning (ML) plays a vital role in addressing climate change by identifying trends, assessing risks, and informing policy \cite{rolnick2022tackling}. Within ML, Natural Language Processing (NLP) has been used to identify climate risks in corporate disclosures, uncovering insights into sustainability efforts \cite{stanny2008corporate}. More recently, large language models (LLMs) have been employed for climate-related tasks, such as supporting public sector decision-making through simulations and modelling \cite{cao2024llm}. LLMs have also been utilised to explore the broader research landscape, especially in mapping innovation trends in countries like the UK \cite{Nesta2021}. By analysing the links between research papers and patents, LLMs help identify areas of scientific work with high commercialisation potential \cite{marx2022reliance}. Furthermore, studies have leveraged LLM-based evaluators for assessing scientific claims and conducting evaluations traditionally requiring human judgement \cite{shankar2024validates, liu2023calibrating, desmond2024evalullm}, as well as combining human and LLM assessments has shown promise in enhancing decision-making processes \cite{yang2024llm}. However, there is a gap in using LLMs to systematically identify climate innovations within a specific region. This study addresses this gap by using LLM evaluators to identify potential climate innovations in the UK, creating a workflow that can be applied to other regions and serve as the foundation for future work.
\section{Methods}
\subsection{Human and LLM evaluation}

In this study, we aimed to identify and evaluate climate-related research papers from six UK-based institutions using both human evaluators and a LLM. The process involved multiple steps, including data collection, abstract filtering, human survey design, and LLM evaluation in different scenarios. Below, we outline each part of the methodology in detail.

\subsection{Data Collection from OpenAlex}

To collect relevant research papers, we utilised the OpenAlex database. OpenAlex \cite{priem2022openalex} is an open, community-curated database designed to provide comprehensive metadata on scholarly publications across a wide range of disciplines. Developed by OurResearch as the successor to the Microsoft Academic Graph (MAG) \cite{wang2020microsoft}, OpenAlex was launched in 2022 with the mission of improving access to scientific knowledge. OpenAlex provides open access to all data through an Application Programming Interface (API), supporting complex queries and offering bulk data downloads, making it suitable for large-scale data analysis projects. In addition to standard interoperable metadata for cross-referencing with other research databases, such as Digital Object Identifiers (DOIs), OpenAlex has applied a comprehensive automated topic classification to all research works, based the Centre for Science and Technology Studies (CWTS) taxonomy. This classifies works according to 4 over-arching domains, 252 subfields and 4516 granular topics and enables a richer analysis of the research corpus. 

Our goal was to focus on climate-related innovations from 6 UK institutions: Imperial College London, University of Leeds, Swansea University, UK Centre for Ecology and Hydrology, University of Cambridge and University of Oxford. We queried OpenAlex for papers where a corresponding author was affiliated with one of these six institutions. We excluded papers with primary topic domain specified as Health Sciences or Social Sciences to maintain a focus on technical and scientific research that is most directly relevant to climate innovation. We also limited works to types associated with primary research (article, pre-print, book chapter or dissertation) and published from year 2000 onwards. The combined query resulted in a dataset of approximately 133,619 works (up to September 2024). The data was downloaded using the \texttt{pyalex} Python Library and \url{https://github.com/J535D165/pyalex}, using the query in \ref{lst:query}.

From this curated dataset, we removed works lacking a listed abstract text or topic classification metadata. Then, having reduced the OpenAlex topic taxonomy to a subset (1860 topic classifications out of 1416 total) we deem most relevant to climate innovation, we removed works with any of the up to four listed topics out of scope. From this final input dataset of \textbf{101,374} works, we randomly selected 95 abstracts for our analysis. Random selection ensured a heterogenous sample of research papers across various disciplines, within the chosen criteria, while avoiding biases that could arise from manual curation or filtering based on specific keywords. Finally, we made the selection of abstracts up to 100 with 5 manually curated abstracts corresponding to research papers from the 6 selected UK-based institutions that led to spin-out climate-tech companies. We 'spiked' the test sample with positive controls in this way in order to enable us to determine the sensitivity of the human- and LLM-based evaluations.

\subsection{Human Evaluation via Survey}\label{sec:humanevaluation}

We implemented a \href{https://imperial.eu.qualtrics.com/jfe/form/SV_0vpoXGvA21UYDnU}{survey} using Qualtrics to collect human evaluations for the 100 selected abstracts. The survey was designed to collect binary responses to seven key questions that assessed different aspects of each paper's potential contribution to climate innovation. The seven questions used in the survey are listed below:

\begin{itemize}

\item \textbf{MITIGATION}: Could this research feasibly lead to a reduction of greenhouse gas emissions or removal of carbon dioxide from the atmosphere?
\item \textbf{TECHNOLOGY}: Does this research describe a technology with practical application?
\item \textbf{READINESS}: Does this research demonstrate that proof-of-concept has been achieved prior to commercialisation or deployment?
\item \textbf{MARKET}: Does a clear commercial market or industry need exist for this research?
\item \textbf{TECH ENABLING}: Rather than a stand-alone technology, does this research represent the fundamental science that might enable future technology development?
\item \textbf{ECO FOCUS}: Was this research conducted with an explicit climate change or sustainability application in mind?
\item \textbf{NEGLECTEDNESS}: Is this research more likely than not to be neglected by existing innovation support mechanisms in the UK?
\end{itemize}

A total of six participants completed the survey, providing binary "Yes" (1) or "No" (0) responses to all seven questions across each of the 100 test abstracts. The binary response format was chosen to simplify analysis and allow for clear comparisons between human and LLM evaluations. Before answering the questions, participants were provided with identical text-based context providing an explanation of each question's purpose and relevance to climate innovation, guiding the decision-making process. This approach was intended to reduce variability in human interpretation and ensure that participants had a similar understanding of the evaluation criteria.

\subsection{Large Language Model (LLM) Evaluation}

In parallel with the human evaluation, we used a large language model (LLM) to analyse the same 100 abstracts and answer the same seven questions mentioned in section \ref{sec:humanevaluation}. To investigate how different levels of context affect the LLM's performance, we set up three distinct scenarios for the LLM prompts (all with temperature = 0 to get almost deterministic answers):

\begin{itemize}
\item \textbf{No-shot}: In this scenario, the LLM was prompted with only the abstract of each paper and a basic instruction: "Do not hallucinate. Only provide truthful answers." This setup aimed to assess the LLM's ability to interpret and evaluate the abstract independently without any additional context or guidance.
   
\item \textbf{Context}: For this scenario, we included the same context that was provided to human evaluators. The LLM received the abstract along with a description of the purpose of each question, similar to what was presented to the survey participants. The goal of this scenario was to determine whether providing context improved the LLM's accuracy and alignment with human judgements.

\item \textbf{Few-Shot Learning}: In the third scenario, we employed a few-shot learning approach to guide the LLM. We supplied the LLM with 10 further example abstracts that described scientific papers leading to spin-out climate-tech companies with high potential to mitigate climate change. The prompt included these examples to demonstrate what types of research could be considered successful climate innovations. The LLM was then asked to evaluate the new abstracts using the provided examples as a reference. This scenario tested whether providing exemplar cases would influence the LLM's responses.
\end{itemize}

In each scenario, the LLM was tasked with generating a "Yes" (1) or "No" (0) answer for each of the seven questions, replicating the human evaluation process. The binary output was designed to facilitate direct comparison with the human responses. The three prompt scenarios were run 6 times to produce a dataset of equivalent size to the human survey.

\subsection{Data Processing and Analysis}

For both human and LLM evaluations, the responses were recorded in a structured dataset. The binary results for each of the seven questions across all 100 abstracts were compiled for subsequent analysis. We aimed to compare the LLM’s responses in each scenario against the human evaluations to identify areas of agreement and discrepancy. Each abstract received seven binary scores from each human evaluator and from the LLM in each of the three scenarios. To quantify general alignment between the human and LLM responses across all respondants and abstracts, we performed pairwise correlation (using the Pearson standard correlation coefficient) between the seven questions and between the datasets. Additionally, we analysed the influence of context and few-shot learning on the LLM's performance. By comparing the LLM's responses across the three scenarios (no context, context, and few-shot), we assessed whether providing additional information or examples improved the accuracy and consistency of the LLM's evaluations relative to the human benchmarks. Finally, we devised a preliminary algorithm to attempt to rank the outputs from the evaluations. First, the \textbf{MITIGATION} question was used as a binary filter to select abstracts identified as broadly relevant to the overall search objective. Secondly, logistic regression was used to determine weightings for the responses to the other six questions when trained on the 5 positive control abstracts. The sum of the weighted scores was used to rank filtered abstracts. Once an optimal LLM scenario was determined, we validated this with a larger test dataset consisting of 990 random abstracts and 10 independent positive control abstracts, as described above. This methodology allows us to systematically evaluate the capability of LLMs in identifying overlooked climate innovations and to explore how different types of context affect their performance. The results from these evaluations are intended to provide insights into how LLMs can be used to supplement human expertise in the search for neglected climate solutions.



\section{Results}

\subsection{Comparison of LLM scenarios against human reasoning }

\begin{figure}[h]
    \centering
    \includegraphics[width = 1\textwidth]{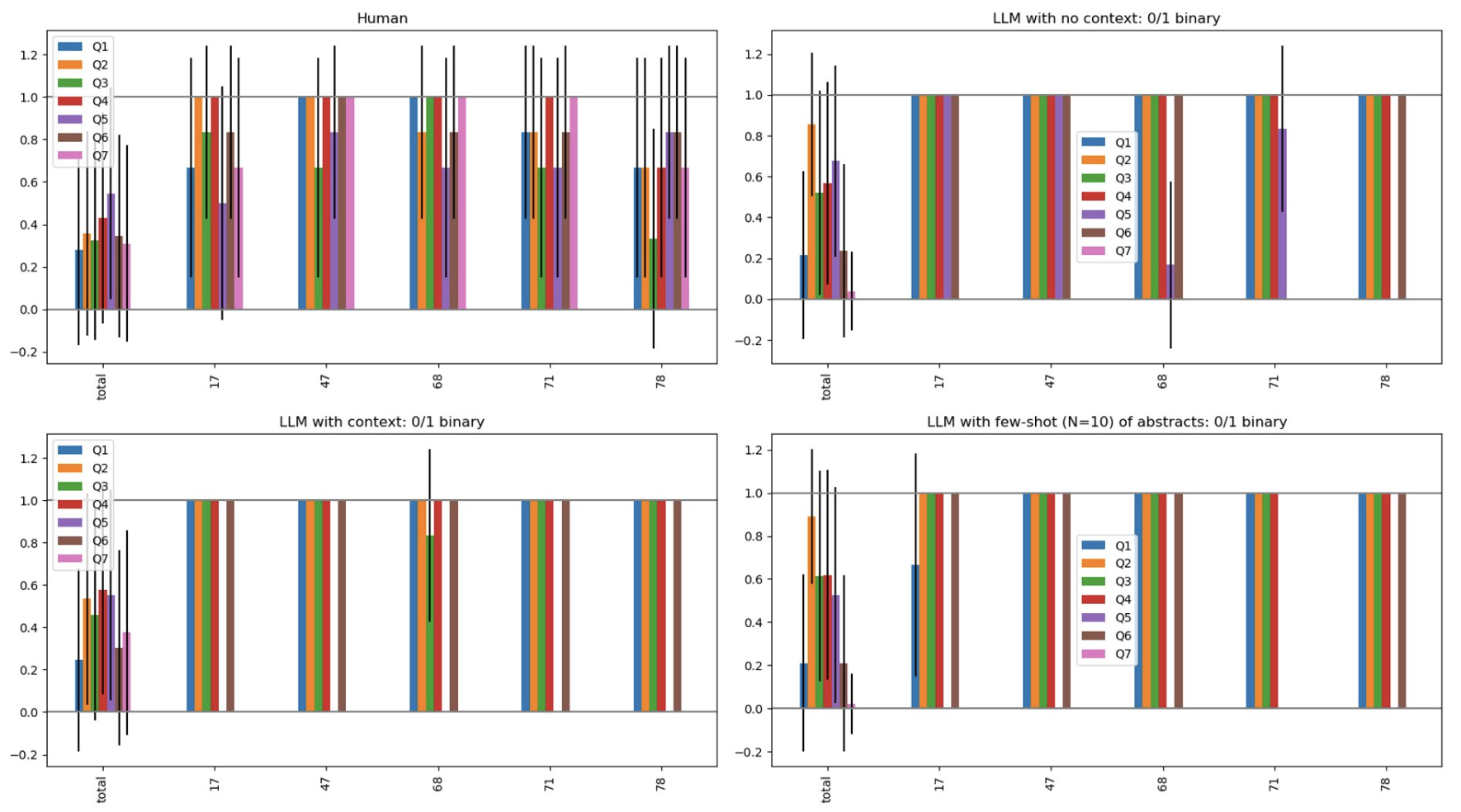}
    \caption{Scoring of positive control abstracts under each scenario}
    \label{fig:scorescenario}
\end{figure}

The 5 positive control abstracts with known linkages to climate tech spin-outs were successfully identified by all three LLM scenarios (no shot, context, few-shot), arguably more convincingly than the 6 human survey participants. Figure\ref{fig:scorescenario} shows the mean respondent scores across the seven questions for the controls (numbered by random id) compared to the mean of all abstracts. In all four datasets, the positive controls generally scored more highly than average. Human scoring was highly variable, compared to predominantly deterministic scoring in the LLM scenarios. Interestingly, while the human respondents generally gave the controls above average scores across all questions, the context and few-shot LLM scenarios scored the controls very low on questions relating to whether the research is fundamentally enabling of technology (Q5) or neglected (Q7), which is predictive of the reality that these technologies have been commercially exploited. Considering the average scores across the 100 abstracts, the LLM provided with the same textual context as the human respondents scored most similarly, particularly clear for technology readiness (Q2) and neglectedness.

\begin{figure}[h]
    \centering
    \includegraphics[width = 0.8\textwidth]{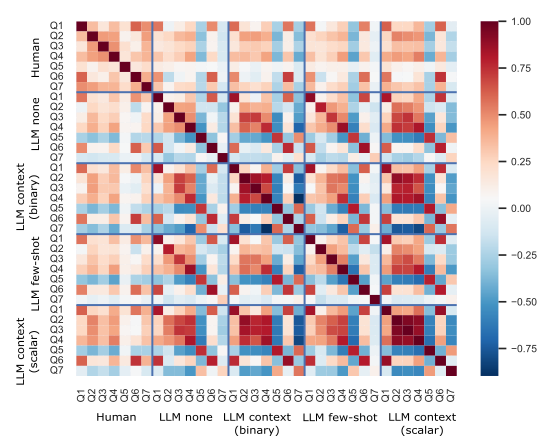}
    \caption{Pearsons coefficient of correlation between questions and scenarios}
    \label{fig:correlation}
\end{figure}

Cohen's Kappa ($\kappa$), a statistical measure that evaluates the level of agreement between two raters beyond what would be expected by chance, was computed for each of the seven questions to compare human responses with those generated by the three LLM scenarios (see Table \ref{tab:cohen_kappa}). The formula for Cohen's Kappa is given by $\kappa = \frac{P_o - P_e}{1 - P_e},
$
where $P_o$ is the observed agreement between the raters, and $P_e$ is the expected agreement by chance. A $\kappa$ value of 1 indicates perfect agreement, 0 indicates agreement equivalent to chance, and negative values suggest disagreement. The results show moderate agreement for questions related to mitigation (Q1) and eco-focus (Q6), particularly under the context scenario ($\kappa = 0.7043$ for Q6). However, there is notably low or even negative agreement for questions concerning technological enabling (Q5) and neglectedness (Q7). The sum of Cohen's Kappa scores across all questions was higest for the context scenario.
\begin{table}[h!]
    \centering
        \caption{Cohen's Kappa scores for each question across different LLM scenarios}
    \begin{tabular}{|l|c|c|c|}
        \hline
        \textbf{Question} & \textbf{Few-shot} & \textbf{Context} & \textbf{No Context} \\ \hline
        \textbf{Q1 (Mitigation)} & 0.5545 & 0.5848 & 0.5473 \\
        \textbf{Q2 (Technology)} & 0.1283 & 0.4151 & 0.1386 \\ 
        \textbf{Q3 (Readiness)} & 0.1914 & 0.2865 & 0.2219 \\ 
        \textbf{Q4 (Market)} & 0.4188 & 0.4156 & 0.4175 \\
        \textbf{Q5 (Tech Enabling)} & -0.0214 & 0.0071 & 0.0317 \\ 
        \textbf{Q6 (Eco Focus)} & 0.5459 & 0.7043 & 0.5945 \\ 
        \textbf{Q7 (Neglectedness)} & -0.0076 & -0.2560 & -0.0630 \\ \hline
    \end{tabular}

    \label{tab:cohen_kappa}
\end{table}
Furthermore, correlation between the seven questions was determined using Pearson's coefficient. Figure~\ref{fig:correlation} shows there is generally weak pairwise correlation between human responses to the survey questions, with the exception of mitigation potential (Q1) and eco-focus (Q6), which correlate best with the LLM responses also. In contrast, the LLM responses reveal a clear binary relationship between two clusters of questions, those suggestive of technical and market readiness for deployment (Q2-4) and those suggestive of early-stage fundamental research with a risk of being neglected (Q5 and Q7). This is most apparent for the LLM scenario where context was provided. Interestingly pairwise correlation between identical questions across datasets is most obvious between the other two LLM scenarios, with either no context or a few-shot of 10 abstracts.
When provided with the same textual context for guidance, these findings suggest that the LLM appears to apply reasoning that is both consistent and the most similar of the scenarios to human scoring.  Human responses vary not only between respondents but between abstracts. Whilst human reasoning draws on nuance and experience, it is evident that both questions and abstracts were judged subjectively with varying outcomes.

\subsection{Evaluation of abstracts identified by LLM}
\subsubsection{Climate change mitigation potential}
\begin{figure}[H]
    \centering
    \begin{subfigure}[t]{0.45\textwidth}
        \centering
        \includegraphics[width=\textwidth]{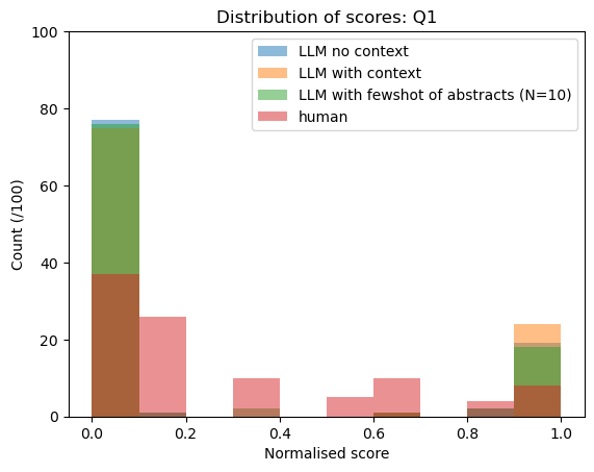}
        \caption{Q1 scores for 100 test abstracts}
        \label{fig:first_subfigure}
    \end{subfigure}
    \hfill 
    \begin{subfigure}[t]{0.45\textwidth}
        \centering
        \includegraphics[width=\textwidth]{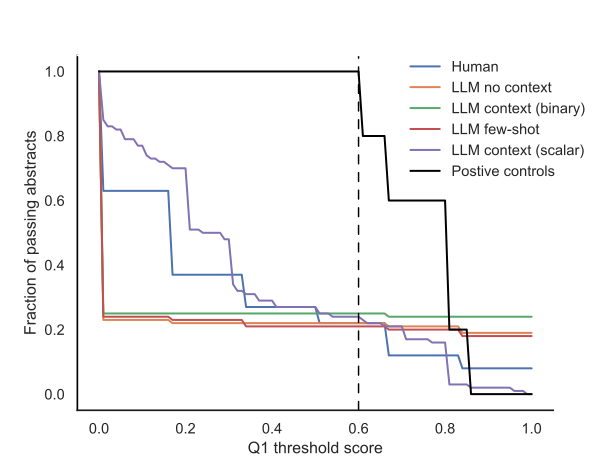}
        \caption{Determining a Q1 pass threshold for filtering abstracts}
        \label{fig:second_subfigure}
    \end{subfigure}
    
    \caption{Mitigation potential of research abstracts (Q1)}
    \label{fig:main_figure}
\end{figure}

Further analysis was performed solely on the LLM with context scenario. Regardless of consistency, the effectiveness of LLM reasoning ultimately depends on a granular assessment of its performance across the unknown test dataset as a whole. Firstly, an objective logic flow was devised for translating scores across the survey questions into a ranking of abstracts. Given the objective focus on climate solutions applications, a primary selection filter was implemented according to the scores for Q1, relating to potential to mitigate climate emissions. Figure~\ref{fig:first_subfigure} shows the distribution of Q1 scores across the four response datasets. The LLM responses are highly deterministic compared to the average of human respondents. A threshold Q1 score of 0.6/1 was found to pass an equivalent fraction of the 100 abstracts across human and LLM responses, including all 5 positive controls. 19 of these passed abstracts were identified in all four scoring scenarios, confirming consistency between human and LLM reasoning on Q1.

\subsubsection{Weighted scoring of Q2-7 determined by Logistic Regression}

To determine the relative predictive quality of the remaining six evaluation questions as indicators for identifying potentially commercialisable climate research papers, we applied logistic regression exclusively to the set of 5 known positive control abstracts related to successful spin-out ventures.

Given a dataset $\mathbf{X}$ consisting of the positive control abstracts, each described by features $\mathbf{X} = \{Q_2, Q_3, Q_4, Q_5, Q_6, Q_7\}$, which correspond to responses to Q2-7, and a target vector $\mathbf{y}$ indicating whether an abstract is part of the positive control set (1) or not (0), logistic regression models the probability that an abstract belongs to the positive control group as:
$P(y = 1 \mid \mathbf{X}) = \frac{1}{1 + \exp(-(\beta_0 + \sum_{i=2}^{7} \beta_i Q_i))}$
where $\beta_0$ is the intercept, and $\beta_i$ are the coefficients (weights) associated with each question response $Q_i$. These coefficients reflect the contribution of each question to the likelihood of an abstract being identified as a potential start-up. Since we only used the positive control abstracts for this regression, the goal was to derive coefficients that best describe the characteristics shared by these successful cases. The coefficients $\beta_i$ were estimated by maximising the likelihood function $L(\beta_0, \beta) = \prod_{j=1}^{N} P(y_j \mid \mathbf{X}_j)^{y_j} \left( 1 - P(y_j \mid \mathbf{X}_j) \right)^{1 - y_j},$
where $N$ is the number of positive control abstracts. To interpret the relative importance of each question, we normalised the coefficients directly so that they sum to 1:$ \hat{w}_i = \frac{\beta_i}{\sum_{j=2}^{7} \beta_j}$. This normalisation process results in a set of positive weights $\hat{w}_i$ that indicate the importance of each question in identifying abstracts with commercialisation potential, as inferred from the positive controls alone. The calculated weights are depicted in Table~\ref{tab:weights}. Applying the calculated weightings to the human and LLM scoring of unknown abstracts, it is clear that the function of deterministic reasoning and a binary scoring presents a challenge for abstract identification. 12 of the 25 abstracts passed by the LLM according to Q1 threshold shared the maximum weighted score, rendering them indistinguishable by means of ranking. This presents a potential problem for the scaling of LLM search to larger abstract datasets, restricting the resolution of the algorithm to approximately only 10\%  and needing subsequent manual assessment.

\begin{table}[t!]
    \centering
    \caption{Weights for Each Question Across Different Models}
    \begin{tabular}{lrrrrr}
        \hline
        & \textbf{Human} & \textbf{Few-shot} & \textbf{Context (binary)} & \textbf{No Context} & \textbf{Context (scalar)} \\
        \hline
        Q2 & 0.212 & 0.164 & 0.211 & 0.035 & 0.663 \\
        Q3 & 0.070 & 0.582 & 0.339 & 0.262 & 0.090 \\
        Q4 & 0.227 & 0.278 & 0.102 & 0.226 & -0.217 \\
        Q5 & 0.015 & -0.544 & -0.235 & 0.172 & -0.082 \\
        Q6 & 0.275 & 0.711 & 0.663 & 0.306 & 0.197 \\
        Q7 & 0.201 & -0.190 & -0.080 & -0.002 & 0.348 \\
        \hline
    \end{tabular}

    \label{tab:weights}
\end{table}

\subsubsection{Extending LLM reasoning with a scalar scoring system }
\begin{figure}[t]
    \centering
    \begin{subfigure}[t]{0.49\textwidth}
        \centering
        \includegraphics[width=\textwidth]{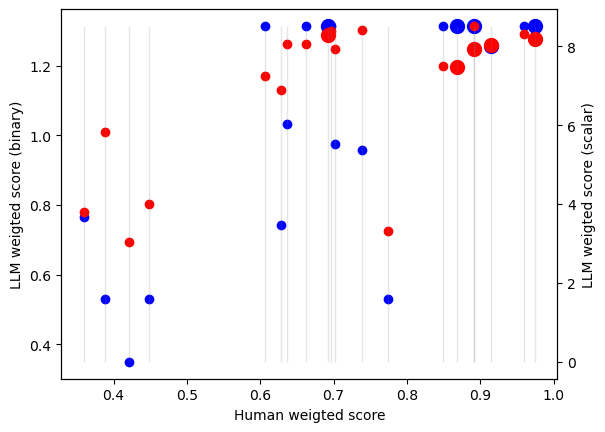}
        \caption{Effect of weighting and scalar (red) scoring on the relationship between overall ranking of selected abstracts. Binary scoring system in blue. Controls are enlarged. }
        \label{fig:scaling_effect}
    \end{subfigure}
    \hfill 
    \begin{subfigure}[t]{0.45\textwidth}
        \centering
        \includegraphics[width=\textwidth]{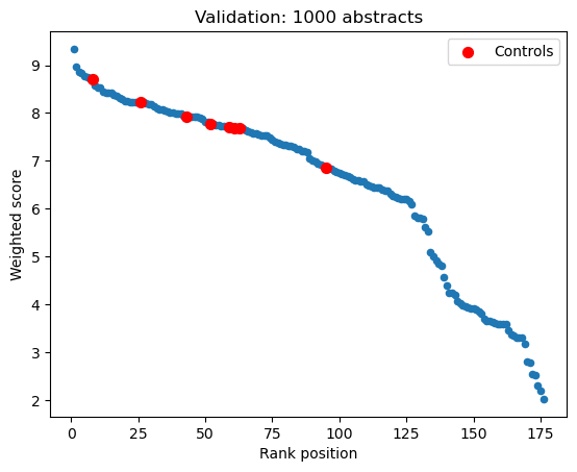}
        \caption{Score ranking of abstracts passing Q1 filter for 1000 abstracts}
        \label{fig:1000_abstracts}
    \end{subfigure}
    
    \caption{Optimisation of test abstract identification by LLM (100 abstracts) and Independent validation of optimised LLM scenario and scoring algorithm for 1000 unseen abstracts}
    \label{fig:100_1000_figure}
\end{figure}

In an attempt to improving the LLM scoring resolution, we reapplied the context approach but with a scalar 1-10 scoring scale rather than a binary system. This type of scoring system was deemed too difficult to apply consistently enough to human reasoning across a large sample of test abstracts. However, the LLM was able to replicate a qualitatively similar scoring outcome but with significantly increased granularity. At the point of Q1 filtering, an almost identical selection of abstracts (24) passed the threshold despite an increased range of possible values (see Figure~\ref{fig:second_subfigure}: LLM context (scalar)). Pearson's correlation coefficient highlights the similarity between the context LLM output in both binary and scalar scoring scenarios (see Figure~\ref{fig:correlation}). Subsequently, with updated weightings calculated (see Table~\ref{tab:weights}: Context (scalar)) and reapplied ranking resolution was improved significantly, with only 2 identical weighted scores out of 24. As depicted in Figure~\ref{fig:scaling_effect} the scaling scoring system (red)  allows differentiation of highly scoring abstracts. Interestingly, updated abstract scores vary both positively and negatively with previous binary scoring system (blue). There is no particular directionality evident.

\subsubsection{Qualitative assessment of research works identified}
The LLM ranked the 5 positive control abstracts at positions 5, 6, 9, 10 and 14, revealing several high scoring candidates among the remaining random abstracts. The top 10 ranked random abstracts were summarised qualitatively by frequency of associated OpenAlex keywords (see Figure~\ref{fig:100_wordcloud} ). Carbon dioxide capture technologies emerge as most frequent by this metric. Further analysis revealed the 10 abstracts are associated with 1 citing patent and 4 climate-tech companies under ownsership by one or more contributing authors.

\subsection{Validation of LLM search tool on larger sample dataset}
The optimised LLM scoring approach (contextualised, weighted and scalar) was validated with a larger dataset of 1000 abstracts, containing a further 10 positive control abstracts associated with spin-outs. 176 papers passed the initial filtering according to potential for climate emissions mitigation (Q1). Only 8 of the 10 positive control abstracts at passed the filter, ranked at positions 8, 26, 43, 52, 59, 61, 63, 95 (see Figure \ref{fig:100_1000_figure}b). Given the low probability that more than half of the filtered papers have as high a potential for commercialisation as the pre-curated controls, this suggests that training the score weightings on more than 5 positive test cases could be worthwhile. 
Keyword assessment of the top 50 ranked abstracts reveal thermal hybrid and photovoltaic technologies as frequently as carbon dioxide capture, observed previously.
\section{Summary and future work}

We show that with appropriate context, prompt engineering and scoring algorithm for interpreting outputs, LLMs represent promising vehicles for the high-throughput identification of high-potential research from the large corpus available. We will further optimise this approach and apply it to the whole UK-wide corpus with the desired output function of predicting climate solutions, a scale of task beyond manual execution. There is significant room for improvement and expansion. Future work will involve experimenting with different LLMs beyond GPT-4o, such as LLaMA or other domain-specific models, to enhance the accuracy and diversity of responses. We will also incorporate Retrieval-Augmented Generation (RAG) to provide the LLMs with a more extensive context, potentially increasing the quality of answers. To accelerate RAG processes, binary quantisation techniques will be applied to reduce computational complexity and improve model efficiency. Additionally, a multi-agent approach will be employed to critically evaluate the LLM’s responses. By using multiple agents, we aim to establish a more robust consensus in the assessment of each abstract. Furthermore, we are developing a relational layer to link papers with their authors, grants, and social media presence. This relational analysis will enable the identification of not only commercially viable papers but also the entrepreneurial potential of the researchers themselves. By understanding authors’ networks and outreach, we hope to target those with a higher likelihood of successfully translating research into impactful climate-related start-up ventures. This study describes the first steps towards the discovery and unearthing of climate solutions in scientific literature.





\clearpage
\bibliographystyle{abbrv}
\bibliography{references}
\clearpage

\appendix
\renewcommand\thefigure{\thesection.\arabic{figure}}    

\section{Appendix}
\setcounter{figure}{0}    

\begin{lstlisting}[language=Python, label={lst:query}, caption={Query using pyalex to extract the academic papers dataset from OpenAlex},captionpos=b]
import pyalex
query = Works().filter(authorships={"institutions":{'country_code':"GB"}}).filter(type='article|preprint|book-chapter|dissertation').filter(authorships={"is_corresponding":'true'}).filter(authorships={"affiliations": {'institution_ids':"https://openalex.org/I82284825|https://openalex.org/I47508984|https://openalex.org/I98677209|https://openalex.org/I130828816|https://openalex.org/I241749|https://openalex.org/I4210092773"}}).filter(publication_year='>1999').filter(primary_topic={"domain": {'id':'!2'}}).filter(primary_topic={"domain": {'id':'!4'}})
\end{lstlisting}

\begin{figure}[H]
    \centering
    \begin{subfigure}[b]{0.49\textwidth}
        \centering
        \includegraphics[width=\textwidth]{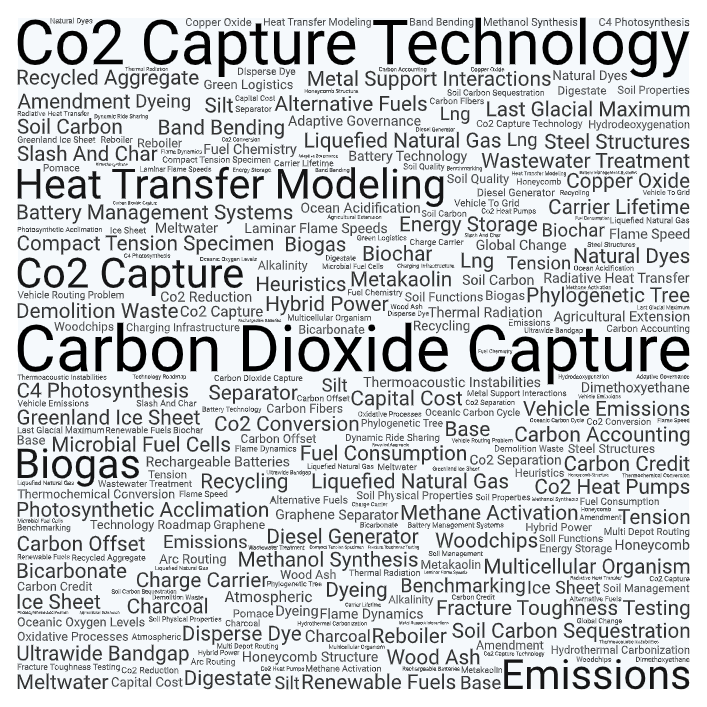}
        \caption{Keyword frequency across 25 abstracts identified by LLM within test dataset (n=100)}
        \label{fig:100_wordcloud}
    \end{subfigure}
    \hfill 
    \begin{subfigure}[b]{0.49\textwidth}
        \centering
        \includegraphics[width=\textwidth]{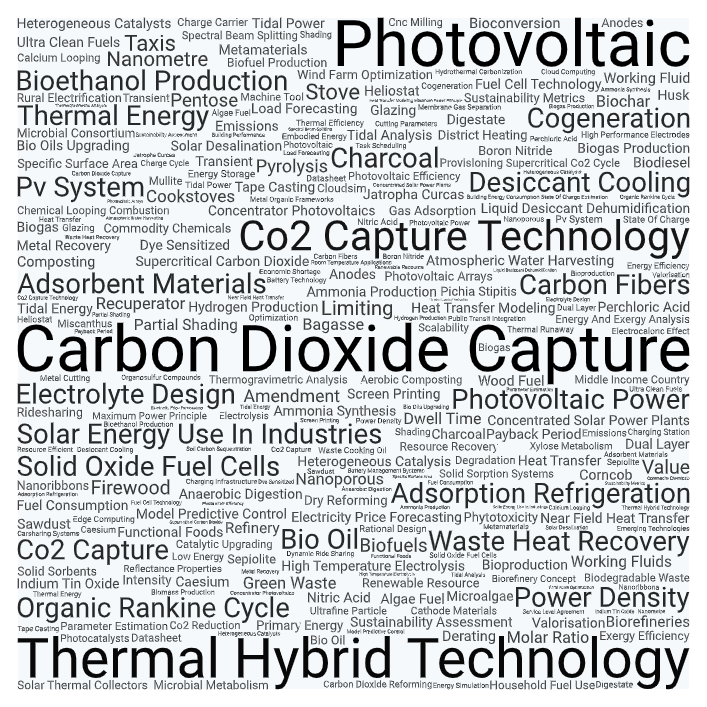}
        \caption{Keyword frequency across top 50 abstracts identified by LLM within validation dataset (n=1000)}
        \label{fig:1000_wordcloud}
    \end{subfigure}
    
    \caption{Keyword frequency in the 100 and 1000 titles and abstracts datasets.}
    \label{fig:1000_figure}
\end{figure}

\end{document}